\documentclass[runningheads]{llncs}
\usepackage[T1]{fontenc}
\usepackage{hyperref}
\usepackage{graphicx}
\usepackage{array}
\usepackage{indentfirst}
\newcommand{\PreserveBackslash}[1]{\let\temp=\\#1\let\\=\temp}
\newcolumntype{C}[1]{>{\PreserveBackslash\centering}p{#1}}
\newcolumntype{R}[1]{>{\PreserveBackslash\raggedleft}p{#1}}
\newcolumntype{L}[1]{>{\PreserveBackslash\raggedright}p{#1}}

\usepackage{pgfplots}
\pgfplotsset{compat=1.18} 

\usepackage{pgfgantt}
\usepackage{makecell}

\begin{document}
\title{Updating Lower and Upper Bounds for the Job-Shop Scheduling Problem Test Instances}
\titlerunning{Updating Lower and Upper Bounds for JSSP Test Instances}
%

\author{Marc-Emmanuel Coupvent des Graviers\inst{1} \and Lotfi Kobrosly\inst{1,2} \and
Christophe Guettier\inst{1} \and Tristan Cazenave \inst{2}}

\authorrunning{ME. Coupvent des Graviers et al.}
%
\institute{Safran Electronics and Defense, France \and
LAMSADE, Université Paris Dauphine-PSL, Place du Maréchal de Lattre de Tassigny, Paris, France}
\maketitle              
\begin{abstract}
The Job-Shop Scheduling Problem (JSSP) and its variant, the Flexible Job-Shop Scheduling Problem (FJSSP), are combinatorial optimization problems studied thoroughly in the literature. Generally, the aim is to reduce the makespan of a scheduling solution corresponding to a problem instance. Thus, finding upper and lower bounds for an optimal makespan enables the assessment of performances for multiple approaches addressed so far. We use OR-Tools, a solver portfolio, to compute new bounds for some open benchmark instances, in order to reduce the gap between upper and lower bounds. We find new numerical lower bounds for multiple benchmark instances, up to closing the Taillard’s \emph{ta33} instance. We also improve upper bounds for four instances, namely Taillard’s \emph{ta26} \& \emph{ta45} and Dauzere's \emph{05a} \& \emph{06a}. Additionally we share an optimal solution for Taillard’s \emph{ta45} as well as Hurink-edata's \emph{car5}.


\end{abstract}
\section{Introduction}

The Job-Shop Scheduling Problem is a staple combinatorial optimization problem studied in the research community. Its study is valuable as it has several applications in the industry \cite{zhang2019review}. The goal of the problem is to order the operations of different jobs on the corresponding machines with the aim of minimizing a certain objective function. Belonging to the NP-hard problem class \cite{watson2003problem,mastrolilli2011hardness}, it has seen several attempts to solve it, such as genetic algorithm, memetic algorithm, branch-and-bound, tabu search, etc. \cite{xiong2022survey}. 
The Flexible Job-Shop Scheduling Problem (FJSSP) is an interesting variation of the JSSP for its relevance to industrial applications ranging from manufacturing \cite{saqlain2023monte} to aeronautics \cite{borreguero2017flexible} and task allocation on computational resources. Those problems are important for industry, as well as operational applications like disaster relief, climate surveillance, and earth monitoring. FJSSP pushes further the NP complexity of JSSP, involving a more combinatorial structure. \cite{pezzella2008genetic,xie2019review}.
In this paper, our aim is to find new lower bounds for test instances of scheduling problems.  The goal is to minimize the scheduling makespan, providing an accurate indicator to which extent the optimization is feasible. In some findings we reach known upper bounds, making further research on the corresponding instances obsolete. We also find new upper bounds for two of Taillard's\cite{taillard1993benchmarks} instances, and the optimal makespan for one of them. Our work is built on OR-Tools, a suite of solvers and a solver-independent interface with several applications \cite{ortools}. Using this software, we model the FJSSP, as well as the JSSP, and run the solvers to provide scheduling solutions, makespans and associated lower bounds.
 
\hyperref[jssp]{Section \ref*{jssp}} gives a detailed presentation of the JSSP and FJSSP as well as some related work. \hyperref[models]{Section \ref*{models}} describes the representations of the FJSSP that we fed to OR-Tools as well as the methodology. Finally, in \hyperref[results]{section \ref*{results}} we present the results of our method on the instances considered.

\section{The Job-Shop Scheduling Problem}
\label{jssp}

\subsection{Problem description}

The JSSP is an NP-hard problem studied broadly throughout the literature. 
The formalized setting of the JSSP can be defined as follows:

\begin{itemize}
    \item[-] A set of $n$ jobs $\mathcal{J}=\{ \mathcal{J}_i ~ | ~ i \in \{ 1..n\}\}$.
    \item[-] A set $\mathcal{M}$ of $m$ machines.
    \item[-] A set $\mathcal{J}_i=\{o_{i,j}, ~ j \in \{1..n_i\}\}$ of $n_i$ operations for every job $i \in \{1..n\}$.
    \item[-] Processing times $d_{j,i}$ for each elementary operation $o_{i,j}$, where $j \in \{1..n_i\}$. 
    \item[-] Operations in a job are ordered and must be processed in said order. More specifically, an operation can only begin when all previous operations of the same job are done.
    \item[-]  Every operation $o_{i,j}$ can be processed on one single machine $m_l$ with $l \in \{1..m\}$. Once processing has started, it cannot be interrupted.
    \item[-] The objective is to optimize a cost function (minimize the makespan, minimize the workload on all machines, etc.) 
\end{itemize}

In this paper, we only consider the minimization of the makespan, which is the total execution time of all operations, as the objective function.
%
Different approaches have tried to model and solve the JSSP. Some schemes are based on Mixed Integer Linear Programming (MILP) \cite{meng2020mixed}, Constraint Programming \cite{zhou1996constraint,da2022industrial}, Tabu Search \cite{nowicki2005advanced} or other local search techniques \cite{beck2011combining}. The emergence of Deep Reinforcement Learning has also had its impact as it considers the JSSP as sequential decision problem \cite{liu2020actor,zhou2020deep}. The usage of Graph Neural Networks has also had a fair amount of consideration used individually or hybridized with other methods, for example with Reinforcement Learning \cite{hameed2020reinforcement} or a randomized $ \epsilon $-greedy approach \cite{abgaryan2024randomized}.

\subsection{The Flexible Job-Shop Scheduling Problem}
\label{fjssp}
The Flexible Job-Shop Scheduling Problem (FJSSP) is a variant of the JSSP where in addition to the available constraints, a subset of operations are no longer related to a unique machine. This means that an operation can be processed by more than one machine, with a different processing time for every compatible machine. To provide a mathematical formulation, we have for every operation $o_{i,j}$ a subset $\mathcal{M}_{i,j}$ of $\mathcal{M}$ with which operation $o_{i,j}$ is compatible. For every couple of compatible operation and machine, denoted $(o_{i,j}, m_l)$, we consider the duration $d_{i,j,l}$ of processing operation $o_{i,j}$ on machine $m_l$. This adds another degree of freedom to the problem and thus another difficulty when solving an instance.
Several representations of the FJSSP and techniques to solve it are reported in the literature, with different objective functions considered \cite{dauzere2024flexible}. They generally compare their results on the benchmark datasets from \cite{behnke2012test,kacem2002pareto,dauzere1995solving,fattahi2007mathematical,barnes1996flexible,hurink1994tabu,brandimarte1993routing}.

\subsection{Review of lower bounds computation}
 
The computation of lower bounds for the FJSSP and the JSSP relies on problem relaxations and extensions as well as mathematical properties from theoretical theorems to computation of longest paths on acyclic graphs \cite{hansmann2014flexible}. An Evolutionary Algorithm \cite{kasapidis2021flexible}, a Mixed-Integer Linear Program  \cite{drotos2009computing}, a Neighborhood Structure Framework \cite{tamassaouet2023general}, and a Failure-Directed Search \cite{heinz4938242reinforcement} were adopted to evaluate and improve these lower bounds. The latest lower bounds values for the datasets we considered are available in \cite{dauzere2024flexible,heinz4938242reinforcement,yuraszeck2023constraint}. 
Specifically for the JSSP lower bounds, we can refer to \emph{optimizer.com} for Demirkol's instances \cite{demirkol1997computational} \footnote{\hyperlink{https://optimizizer.com/DMU.php}{https://optimizizer.com/DMU.php}} and Taillard's instances \cite{taillard1993benchmarks}, with  corresponding references \footnote{\hyperlink{https://optimizizer.com/TA.php}{https://optimizizer.com/TA.php}}.

\section{Implementing the Job-Shop Scheduling Solver}
\label{models}

The vanilla model of the FJSSP, which can also represent the JSSP, is adapted from \cite{echeverria2024leveraging} and is explained in \hyperref[fjssp_cp_model]{Model}. It is considered as a Constraint Programming (CP) problem and uses built-in functions of OR-Tools. The model finds tasks' starting and ending timestamps without explicitly looking for task order. It also chooses task to machine allocation by explicitly requiring every task to be allocated to exactly one machine. It forbids overlap between tasks interval (from starting to ending timestamp) allocated to the same machine. Lastly, it constrains tasks pertaining to the same job to be scheduled according to their number in the job.

\begin{table}
    \centering
    \begin{tabular}{L{2cm} L{8cm}}
        \hline
        \textbf{Model} & FJSSP CP model \\
        \hline
        Minimize: & \space makespan \\
         & \\
        S.t: & \space $\forall i,j: End_{i,j} \leq makespan$ \\
         & \space  $A_{i,j,l} \rightarrow I_{i,j,l}$ \\
         & \space $I_{i,j,l} \rightarrow Start_{i,j} + d_{i,j,l} = End_{i,j} $ \\
         & \space $\forall l \in \mathcal{M}: \verb|NoOverlap|_{i,j} (I_{i,j,l}) $ \\
         & \space $\forall i,j: \verb|ExactlyOne|_{l \in \mathcal{M}} (A_{i,j,l})$ \\
         & \space $\forall i, \forall j,k>j: Start_{i,k} \geq End_{i,j} $ \\
        \hline
         & 
    \end{tabular}
    \label{fjssp_cp_model}
    \caption{CP model description where: $Start_{i,j}$ and $End_{i,j}$ are the starting and ending execution times of task $o_{i,j}$. $A_{i,j,l}$ is the machine allocation for said operation, whose value is \textbf{True} if task $o_{i,j}$ is scheduled on machine $l$. $I_{i,j,l}$ is an optional interval for task $o_{i,j}$ when scheduled on machine $l$ and $d_{i,j,l}$ is the corresponding duration. Variable types and solver configuration are described in §3.}
\end{table}

%
We are using OR-tools solver to search for minimum makespan and keep the best lower and upper bounds. We use an integer (\verb+NewIntVar+) for every task start and end timestamps, regardless of machine allocation. Every timestamp is initially bounded by the sum of tasks durations. Then we use an optional integer interval (\verb+NewOptionalIntervalVar+) for every combination of task and machine. Global constraint \verb+AddNoOverlap+ is then applied per machine, guaranteeing disjunction between every optional interval compatible with the given machine. Some specific OR-Tools parameters have been tuned by iterative tests to improve lower bound search time:
\begin{itemize}
    \item[] \verb+use_precedences_in_disjunctive_constraint = False+
    \item[] \verb+use_disjunctive_constraint_in_cumulative = False+
    \item[] \verb+num_search_workers = 48+
\end{itemize}

The solver is OR-tools 9.9.3963, running on AMD EPYC 7763 CPU embedded in an AI server. Solving time was initially bound to 900s as in \cite{heinz4938242reinforcement} even if the thread number is not similar. However, the lower bound was still improving on some instances after this initial 900s time out, as depicted for example in Figure \ref{dauzere_16a}. Solving time was later extended up to 50000s. We found that solving performances were highly dependant on the CPU power and search workers' number, probably due to automated subsolver selecting from the internal OR-Tools portfolio. However, systematic sensitivity analysis is not the purpose of this paper.

\section{Results}
\label{results}

We show in this section lower and upper bounds improvement found by our experiments on the instances of the FJSSP Dauzere \cite{dauzere1995solving}, and Hurink \cite{hurink1994tabu}, as well as those of the JSSP labelled TA \cite{taillard1993benchmarks} and Demirkol \cite{demirkol1997computational}. Other benchmark instances of the FJSSP \cite{barnes1996flexible,kacem2002pareto,behnke2012test,fattahi2007mathematical,brandimarte1993routing}, have already been solved \cite{dauzere2024flexible} (except for Mk10 instance from \cite{brandimarte1993routing}).
We are using dataset instances stored in \footnote{\href{https://github.com/SchedulingLab/fjsp-instances}{github.com/SchedulingLab/fjsp-instances}}

We used two methods to find new lower bounds. The first method consists in searching for minimum makespan as described previously in FJSSP model and storing the best lower bound found by OR-Tools. The second method checks for solution feasibility while constraining the makespan below a given threshold, which is increased incrementally outside of OR-Tools. We call these methods OPT and SAT respectively in result tables, along with experiments timeout values. Our result tables only reports instances for which we improve either the upper or lower bound. Lack of improvement for a given method or timeout is marked with ’-’ in the result tables. We also show Gantt's diagrams for the optimal solutions we found for \emph{ta45} instance from Taillard's \cite{taillard1993benchmarks} in figure 1 and for \emph{car5} instance from Hurink's \emph{vdata} \cite{hurink1994tabu} in figure 2.

\begin{table}[h!]
    \centering
    \label{brandimarte_results}
    \begin{tabular}{| C{3cm} | C{1cm} | C{1cm} | C{1.3cm} | C{1.3cm}| C{1.3cm} | C{2.2cm} |}
        \hline
        \textbf{} & \multicolumn{2}{|c|}{known bounds} & \multicolumn{3}{|c|}{new Lower Bound} & \multicolumn{1}{|c|}{} \\         
        \hline
        \multicolumn{1}{|c|}{Instance} & \multicolumn{1}{|c|}{LB} & \multicolumn{1}{|c|}{UB} & \multicolumn{1}{|c|}{OPT-900s} & \multicolumn{1}{|c|}{SAT-900s} & \multicolumn{1}{|c|}{OPT-50000s}& \multicolumn{1}{c|}{Gap reduction}\\
        \hline
            mk10 & 189 & 193 & \textbf{190} & - & - & 25\% \\
        \hline
    \end{tabular}
    \caption{Results on Brandimarte FJSSP dataset \cite{brandimarte1993routing}.}
\end{table}

\begin{table}[h!]
    \centering
    \label{DMU_results}
    \begin{tabular}{| C{3cm} | C{1cm} | C{1cm} | C{1.3cm} | C{1.3cm}| C{1.3cm} | C{2.2cm} |}
        \hline
        \textbf{} & \multicolumn{2}{|c|}{known bounds} & \multicolumn{3}{|c|}{new Lower Bound} & \multicolumn{1}{|c|}{} \\         
        \hline
        \multicolumn{1}{|c|}{Instance} & \multicolumn{1}{|c|}{LB} & \multicolumn{1}{|c|}{UB} & \multicolumn{1}{|c|}{OPT-900s} & \multicolumn{1}{|c|}{SAT-900s} & \multicolumn{1}{|c|}{OPT-50000s}& \multicolumn{1}{c|}{Gap reduction}\\
        \hline        
            cscmax\_20\_15\_10 & 3130 & 3248 & - & - & \textbf{3139} & 7\% \\
            cscmax\_30\_15\_2 & 4042 & 4156 & - & - & \textbf{4053} & 9\% \\
            cscmax\_30\_15\_9 & 4170 & 4297 & - & - & \textbf{4179} & 7\% \\
            cscmax\_40\_15\_3 & 5014 & 5169 & - & - & \textbf{5016} & 1\% \\
            cscmax\_40\_15\_7 & 5121 & 5173 & - & - & \textbf{5122} & 2\% \\
            cscmax\_40\_20\_5 & 5500 & 5688 & - & - & \textbf{5502} & 1\% \\
            cscmax\_40\_20\_6 & 5650 & 5779 & - & - & \textbf{5652} & 1.5\% \\
            cscmax\_40\_20\_9 & 5619 & 5868 & - & - & \textbf{5621} & 1\% \\
            cscmax\_50\_15\_3 & 6046 & 6189 & 6120 & - & \textbf{6123} & 54\% \\
            cscmax\_50\_15\_4 & 6128 & 6196 & \textbf{6168} & - & - & 59\% \\
            cscmax\_50\_15\_6 & 6430 & 6463 & - & - & \textbf{6434} & 12\% \\
            cscmax\_50\_15\_8 & 6103 & 6207 & 6123 & 6124 & \textbf{6127} & 23\% \\
            cscmax\_50\_15\_10 & 6013 & 6136 & 6106 & - & \textbf{6107} & 76\% \\  
            cscmax\_50\_20\_1 & 6463 & 6718 & - & - & \textbf{6468} & 2\% \\
            cscmax\_50\_20\_4 & 6517 & 6747 & - & - & \textbf{6518} & 0,4\% \\
            rcmax\_20\_15\_8 & 2638 & 2669 & - & - & \textbf{2641} & 10\% \\
            rcmax\_20\_20\_4 & 2995 & 3046 & - & - & \textbf{3014} & 37\% \\
            rcmax\_20\_20\_6 & 3188 & 3244 & - & - & \textbf{3207} & 34\% \\
            rcmax\_30\_15\_9 & 3396 & 3430 & - & - & \textbf{3397} & 3\% \\
            rcmax\_30\_15\_10 & 3418 & 3492 & \textbf{3481} & - & - & 85\% \\
            rcmax\_30\_20\_10 & 3717 & 3812 & - & - & \textbf{3723} & 6\% \\
        \hline
    \end{tabular}
    \caption{Results on Demirkol JSSP dataset \cite{demirkol1997computational}.}
\end{table}

\begin{table}[h!]
    \centering
    \label{hurink_rdata_results}
    \begin{tabular}{| C{3cm} | C{1cm} | C{1cm} | C{1.3cm} | C{1.3cm}| C{1.3cm} | C{2.2cm} |}
        \hline
        \textbf{} & \multicolumn{2}{|c|}{known bounds} & \multicolumn{3}{|c|}{new Lower Bound} & \multicolumn{1}{|c|}{} \\         
        \hline
        \multicolumn{1}{|c|}{Instance} & \multicolumn{1}{|c|}{LB} & \multicolumn{1}{|c|}{UB} & \multicolumn{1}{|c|}{OPT-900s} & \multicolumn{1}{|c|}{SAT-900s} & \multicolumn{1}{|c|}{OPT-50000s}& \multicolumn{1}{c|}{Gap reduction}\\
        \hline
            abz7 & 493 & 522 & 496 & \textbf{497} & - & 13\% \\
            abz8 & 507 & 535 & 508 & \textbf{509} & - & 7\% \\
            la21 & 808 & 825 & - & - & \textbf{809} & 5\% \\
            la22 & 741 & 753 & - & - & \textbf{745} & 33\% \\
            la23 & 816 & 831 & 819 & - & \textbf{820} & 26\% \\
            la24 & 775 & 795 & 778 & - & \textbf{780} & 25\% \\
            la25 & 768 & 779 & - & - & \textbf{771} & 27\% \\
        \hline
    \end{tabular}
    \caption{Results on Hurink rdata FJSSP dataset \cite{hurink1994tabu}.}
\end{table}

\begin{table}[h!]
    \centering
    \label{taillard_results}
    \begin{tabular}{| C{3cm} | C{1cm} | C{1cm} | C{1.3cm} | C{1.3cm}| C{1.3cm} |  C{1.3cm} | C{2.2cm} |}
        \hline
        \textbf{} & \multicolumn{2}{|c|}{known bounds} & \multicolumn{3}{|c|}{new Lower Bound} & new UB & \multicolumn{1}{|c|}{} \\         
        \hline
        \multicolumn{1}{|c|}{Instance} & \multicolumn{1}{|c|}{LB} & \multicolumn{1}{|c|}{UB} & \multicolumn{1}{|c|}{OPT-900s} & \multicolumn{1}{|c|}{OPT-3600s} & \multicolumn{1}{|c|}{OPT-50000s}& & \multicolumn{1}{c|}{Gap reduction}\\
        \hline 	
            Ta18 & 1382 & 1396 & - & - & \textbf{1393} & - & 78\% \\
            Ta22 & 1576 & 1600 & - & - & \textbf{1581} & - & 20\% \\
            Ta23 & 1533 & 1557 & - & - & \textbf{1544} & - & 45\% \\
            Ta25 & 1575 & 1595 & - & - & \textbf{1586} & - & 55\% \\
            Ta26 & 1609 & 1645 & - & - & \textbf{1616} &  \textbf{1643} & 25\% \\
            Ta27 & 1665 & 1680 & - & - & \textbf{1673} & - & 53\% \\
            Ta29 & 1597 & 1625 & - & - & \textbf{1607} & - & 35\% \\
            Ta30 & 1539 & 1584 & - & - & \textbf{1552} & - & 28\% \\
            Ta33 & 1790 & 1791 & - & \textbf{1791} &  & &\textbf{100\%} \\
            Ta40 & 1652 & 1670 & - & - & \textbf{1653} & - & 5\% \\
            Ta41 & 1912 & 2006 & - & - & \textbf{1915} & - & 3\% \\
            Ta42 & 1887 & 1939 & - & - & \textbf{1888} & - & 2\% \\
            Ta45 & 1997 & 2000 & - & - & - &  \textbf{1997} & \textbf{100\%} \\
            Ta46 & 1966 & 2006 & - & - & \textbf{1967} & - & 2,5\% \\
            Ta47 & 1816 & 1889 & - & - & \textbf{1819} & - & 4\% \\
        \hline
    \end{tabular}
    \caption{Results on Taillard JSSP dataset \cite{taillard1993benchmarks}.}
\end{table}

\begin{table}[h!]
    \centering
    \label{dauzere_results}
    \begin{tabular}{| C{3cm} | C{1cm} | C{1cm} | C{1.3cm} | C{1.3cm}| C{1.3cm} |  C{1.3cm} | C{2.2cm} |}
        \hline
        \textbf{} & \multicolumn{2}{|c|}{known bounds} & \multicolumn{3}{|c|}{new Lower Bound} & new UB & \multicolumn{1}{|c|}{} \\         
        \hline
        \multicolumn{1}{|c|}{Instance} & \multicolumn{1}{|c|}{LB} & \multicolumn{1}{|c|}{UB} & \multicolumn{1}{|c|}{OPT-900s} & \multicolumn{1}{|c|}{OPT-3600s} & \multicolumn{1}{|c|}{OPT-50000s}& & \multicolumn{1}{c|}{Gap reduction}\\
        \hline 	
            05a & 2192 & 2203 & 2193 & - & \textbf{2195'} & \textbf{2199'} & 64\% \\
            06a & 2163 & 2171 & - & - & \textbf{2164} & \textbf{2169'} & 37\% \\
            11a & 2018 & 2037 & \textbf{2019} & - & - & - & 5\% \\
            13a & 2197 & 2236 & 2201 & 2202 & \textbf{2206} & - & 23\% \\
            16a & 2193 & 2231 & 2194 & 2198 & \textbf{2202} & - & 24\% \\
            17a & 2088 & 2105 & - & \textbf{2089} & - & - & 5\% \\
        \hline
    \end{tabular}
    \caption{Results on Dauzere FJSSP dataset \cite{dauzere1995solving}. Specifically instance 05a was solved longer than 5000 seconds. Lower bound 2195 was found after 180000s, and upper bound 2199 was found after 145000s. We also had to solve instance 06a a second time to confirm the new upper bound, which took 160000s}
\end{table}

\begin{table}[h!]
    \centering
    \label{hurink_vdata_results}
    \begin{tabular}{| C{3cm} | C{1cm} | C{1cm} | C{1.3cm} | C{1.3cm}| C{1.3cm} |  C{1.3cm} | C{2.2cm} |}
        \hline
        \textbf{} & \multicolumn{2}{|c|}{known bounds} & \multicolumn{3}{|c|}{new Lower Bound} & new UB & \multicolumn{1}{|c|}{} \\         
        \hline
        \multicolumn{1}{|c|}{Instance} & \multicolumn{1}{|c|}{LB} & \multicolumn{1}{|c|}{UB} & \multicolumn{1}{|c|}{OPT-900s} & \multicolumn{1}{|c|}{OPT-3600s} & \multicolumn{1}{|c|}{OPT-50000s}& & \multicolumn{1}{c|}{Gap reduction}\\
        \hline 	
            car5 & 4909 & 4910 & - & - & - & \textbf{4909} & \textbf{100\%} \\
        \hline
    \end{tabular}
    \caption{Results on Hurink vdata FJSSP dataset \cite{hurink1994tabu}.}
\end{table}

\clearpage

\begin{figure}[h!]
\raggedleft
\begin{tikzpicture}

\definecolor{darkgray176}{RGB}{176,176,176}

    \begin{axis}[
        width=1.1\textwidth,
        height=4.5cm,
        tick align=outside,
        tick pos=left,
        title={},
        x grid style={darkgray176},
        xmin=-0.5, xmax=4908.5,
        xlabel={Time},
        xtick style={draw=none},
        y dir=reverse,
        y grid style={darkgray176},
        ymin=-0.5, ymax=5.5,
        ytick style={draw=none},
        ylabel={Machine}
    ]
        \addplot graphics [includegraphics cmd=\pgfimage,xmin=-0.5, xmax=4908.5, ymin=5.5, ymax=-0.5] {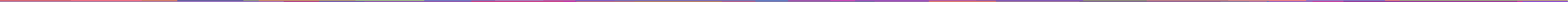};
    \end{axis}

\end{tikzpicture}
\label{vdata_car5_figure}
\caption{Gantt's diagram of the optimal solution for the instance \emph{car5} of Hurink's \emph{vdata} \cite{hurink1994tabu}. Operations of the same colour belong to the same job, and white spaces represent inactivity on a machine. For this specific instance machines are used continuously without interleave between tasks.}

\vspace{2cm}

\begin{tikzpicture}[trim axis left]
    
    \definecolor{darkgray176}{RGB}{176,176,176}
    
    \begin{axis}[
        width=1.3\textwidth,
        height=10cm,
        tick align=outside,
        tick pos=left,
        title={},
        x grid style={darkgray176},
        xmin=-0.5, xmax=1996.5,
        xlabel={Time},
        xtick style={draw=none},
        y dir=reverse,
        y grid style={darkgray176},
        ymin=-0.5, ymax=19.5,
        ytick style={draw=none},
        ylabel={Machine}
    ]
        \addplot graphics [includegraphics cmd=\pgfimage,xmin=-0.5, xmax=1996.5, ymin=19.5, ymax=-0.5] {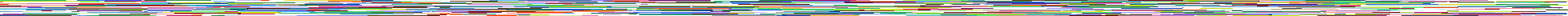};
    \end{axis}
    
\end{tikzpicture}
\caption{Gantt's diagram of the optimal solution for the instance \emph{ta45} of Taillard's \cite{taillard1993benchmarks}. Operations of the same colour belong to the same job, and white spaces represent inactivity on a machine. There are multiple remaining interleaves in machine usage. The makespan is thus constrained by job structure.}
\end{figure}

\clearpage

\begin{figure}[h!]
\centering
\begin{tikzpicture}
\begin{axis} [
    xmode=log,
    xlabel={\texttt{Solving time}},
    ylabel={\texttt{Dauzere \emph{16a} lower bound}},
    ]
\addplot table [x=a, y=b, col sep=comma] {data.csv};
\addplot[
    color=red,
    mark=*,
    only marks,
    mark options={fill=red},
] coordinates {(450.52, 2193)};
\end{axis}
\end{tikzpicture}
\caption{Lower bound for Dauzere \cite{dauzere1995solving} instance \emph{16a}, along solving time (log scale in seconds). Previously best known lower bound is 2193 in red. This previously known value is in the regime area where adding computational time is sufficient to find new values.}
\label{dauzere_16a}
\end{figure}
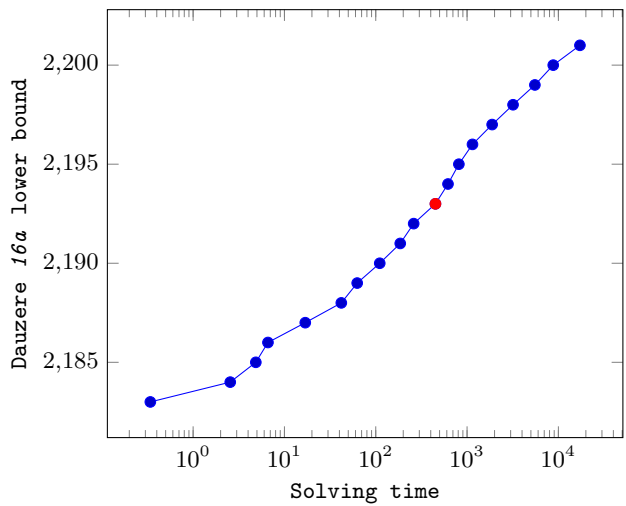

\section{Conclusion}
 
We used a simple CP model, OrTools and a multicore industrial computer to assess the performance of this solution over a set of FJSSP benchmarks. We have surprisingly found better lower bounds than previously known values, and chose to share them. Our approach does not provides proof on those new lower bounds as the risk of OrTools implementation error cannot be formally eliminated. It is worth mentioning however that we have not found inconsistent values such as LB higher than known UB, or a solution lower than known LB.

We have also found five new upper bounds with the associated solutions. These solutions are directly verified for correctness, without relying on solver correctness. While we do not provide a new solving approach, we believe it could help the research community to share these new bounds to better assess the performance of other search algorithms.

%
%
%
\clearpage

\small{
\bibliographystyle{splncs04}
\bibliography{bibliography}
}

\clearpage

\section*{Appendix}

This annexes provides solution for improved upper bounds, and summarizes up to date lower and upper bounds for multiple datasets.

\begin{table}[h!]
\centering
\begin{tabular}{|c|>{\centering\arraybackslash}p{10cm}|>{\centering\arraybackslash}p{1.5cm}|}
\hline
\textbf{Machine} & \textbf{Job order} & \textbf{end time} \\
\hline
1 & \makecell{ $7_{0}$,$7_{1}$,$5_{1}$,$7_{3}$,$4_{2}$,$3_{3}$,$9_{2}$,$7_{6}$,$7_{7}$,$7_{8}$,$7_{9}$,$1_{3}$,$9_{4}$,$1_{4}$,$2_{7}$,$1_{6}$,$7_{13}$,$5_{11}$,$1_{7}$,$9_{9}$,$1_{9}$,$2_{11}$,$4_{12}$, \\
$8_{12}$,$6_{7}$,$6_{8}$,$9_{13}$,$9_{14}$,$3_{11}$,$6_{11}$,$9_{15}$,$4_{17}$,$8_{18}$,$1_{17}$,$7_{22}$,$4_{19}$,$6_{15}$,$9_{18}$,$4_{20}$ } &  2169 \\
\hline
2 & \makecell{ $5_{0}$,$8_{1}$,$9_{1}$,$2_{1}$,$0_{0}$,$5_{3}$,$4_{4}$,$5_{5}$,$5_{6}$,$2_{4}$,$8_{5}$,$3_{6}$,$2_{6}$,$5_{9}$,$3_{7}$,$4_{10}$,$8_{8}$,$7_{14}$,$3_{8}$,$1_{8}$,$5_{13}$,$5_{14}$,$2_{12}$, \\
$7_{16}$,$7_{17}$,$9_{11}$,$0_{7}$,$1_{13}$,$2_{14}$,$1_{14}$,$4_{15}$,$8_{15}$,$8_{16}$,$4_{16}$,$7_{20}$,$2_{19}$,$2_{20}$,$8_{19}$,$3_{14}$,$3_{15}$,$8_{21}$,$1_{19}$ } &  2169 \\
\hline
3 & \makecell{ $9_{0}$,$1_{0}$,$4_{1}$,$8_{2}$,$7_{4}$,$4_{3}$,$7_{5}$,$2_{3}$,$0_{1}$,$4_{6}$,$5_{8}$,$9_{5}$,$9_{6}$,$7_{12}$,$0_{2}$,$2_{8}$,$6_{4}$,$6_{5}$,$8_{9}$,$9_{10}$,$1_{10}$,$1_{11}$,$3_{9}$, \\
$9_{12}$,$8_{13}$,$4_{14}$,$8_{14}$,$7_{18}$,$7_{19}$,$0_{10}$,$2_{18}$,$9_{16}$,$6_{13}$,$6_{14}$,$1_{18}$,$3_{16}$,$3_{17}$ } &  2169 \\
\hline
4 & \makecell{ $4_{0}$,$2_{0}$,$6_{0}$,$5_{2}$,$2_{2}$,$1_{1}$,$8_{3}$,$8_{4}$,$1_{2}$,$9_{3}$,$6_{2}$,$8_{6}$,$4_{7}$,$6_{3}$,$1_{5}$,$4_{9}$,$8_{7}$,$2_{9}$,$0_{3}$,$5_{12}$,$4_{11}$,$7_{15}$,$0_{5}$,$6_{6}$, \\
$1_{12}$,$0_{8}$,$6_{9}$,$2_{16}$,$6_{10}$,$2_{17}$,$3_{12}$,$3_{13}$,$7_{21}$,$2_{21}$,$8_{20}$,$2_{22}$,$0_{14}$ } &  2169 \\
\hline
5 & \makecell{ $8_{0}$,$3_{0}$,$7_{2}$,$3_{1}$,$3_{2}$,$6_{1}$,$5_{4}$,$3_{4}$,$4_{5}$,$3_{5}$,$5_{7}$,$2_{5}$,$7_{10}$,$7_{11}$,$4_{8}$,$5_{10}$,$9_{7}$,$9_{8}$,$2_{10}$,$0_{4}$,$8_{10}$,$8_{11}$,$0_{6}$, \\
$2_{13}$,$4_{13}$,$2_{15}$,$0_{9}$,$3_{10}$,$1_{15}$,$1_{16}$,$8_{17}$,$6_{12}$,$0_{11}$,$4_{18}$,$0_{12}$,$9_{17}$,$0_{13}$,$7_{23}$,$7_{24}$,$6_{16}$,$8_{22}$ } &  2169 \\
\hline
\end{tabular}
\caption{New Upper Bound solution of Dauzere : 06a. This table provides job and task order and last timestamp for every machine. Job and tasks are formalized as $Job_{Task}$. Solving time is 160000 seconds.}
\end{table}

\begin{table}[h!]
\centering
\begin{tabular}{|c|>{\centering\arraybackslash}p{10cm}|>{\centering\arraybackslash}p{1.5cm}|}
\hline
\textbf{Machine} & \textbf{Job order} & \textbf{end time} \\
\hline
1 & \makecell{ $3_{0}$,$9_{1}$,$9_{2}$,$4_{1}$,$8_{4}$,$5_{1}$,$1_{2}$,$0_{4}$,$8_{7}$,$2_{5}$,$2_{6}$,$9_{8}$,$4_{6}$,$0_{7}$,$7_{7}$,$7_{8}$,$9_{10}$,$4_{9}$,$6_{4}$,$2_{10}$,$4_{11}$,$3_{6}$,$3_{7}$, \\ 
$2_{13}$,$4_{14}$,$7_{14}$,$3_{10}$,$6_{11}$,$9_{16}$,$3_{12}$,$8_{20}$,$5_{11}$,$9_{18}$,$2_{18}$,$1_{17}$,$1_{18}$,$3_{16}$,$7_{24}$,$6_{15}$,$1_{19}$ } &  2199 \\
\hline
2 & \makecell{ $4_{0}$,$5_{0}$,$8_{3}$,$7_{1}$,$3_{3}$,$8_{5}$,$2_{3}$,$8_{6}$,$9_{7}$,$5_{4}$,$7_{5}$,$1_{4}$,$5_{8}$,$4_{7}$,$0_{8}$,$8_{11}$,$8_{12}$,$4_{10}$,$9_{12}$,$9_{13}$,$1_{10}$,$1_{11}$, \\ 
$0_{11}$,$3_{8}$,$8_{16}$,$8_{17}$,$3_{11}$,$7_{17}$,$7_{18}$,$7_{19}$,$9_{17}$,$7_{20}$,$5_{12}$,$7_{22}$,$8_{21}$,$5_{14}$ } &  2197 \\
\hline
3 & \makecell{ $8_{0}$,$8_{1}$,$8_{2}$,$0_{0}$,$0_{1}$,$7_{2}$,$9_{4}$,$9_{5}$,$9_{6}$,$2_{4}$,$4_{3}$,$1_{3}$,$4_{5}$,$7_{6}$,$1_{5}$,$8_{10}$,$4_{8}$,$2_{8}$,$0_{9}$,$2_{9}$,$1_{8}$,$6_{5}$,$8_{14}$,$4_{12}$, \\ 
$9_{14}$,$8_{15}$,$7_{13}$,$3_{9}$,$9_{15}$,$8_{18}$,$7_{16}$,$1_{14}$,$2_{16}$,$5_{10}$,$0_{13}$,$3_{14}$,$3_{15}$,$2_{20}$,$2_{21}$,$4_{20}$,$3_{17}$,$6_{16}$ } &  2199 \\
\hline
4 & \makecell{ $9_{0}$,$2_{0}$,$7_{0}$,$3_{2}$,$1_{1}$,$0_{3}$,$7_{3}$,$5_{2}$,$5_{3}$,$3_{4}$,$5_{5}$,$5_{6}$,$5_{7}$,$6_{1}$,$2_{7}$,$1_{6}$,$6_{2}$,$1_{7}$,$3_{5}$,$8_{13}$,$7_{10}$,$2_{11}$,$7_{11}$, \\ 
$6_{8}$,$6_{9}$,$1_{12}$,$1_{13}$,$2_{14}$,$7_{15}$,$2_{15}$,$8_{19}$,$6_{12}$,$2_{17}$,$1_{16}$,$6_{13}$,$2_{19}$,$6_{14}$,$2_{22}$,$8_{22}$ } &  2199 \\
\hline
5 & \makecell{ $6_{0}$,$1_{0}$,$3_{1}$,$2_{1}$,$9_{3}$,$0_{2}$,$2_{2}$,$4_{2}$,$7_{4}$,$0_{5}$,$4_{4}$,$0_{6}$,$8_{8}$,$8_{9}$,$9_{9}$,$5_{9}$,$6_{3}$,$7_{9}$,$9_{11}$,$0_{10}$,$1_{9}$,$6_{6}$,$6_{7}$,$2_{12}$ \\ 
,$7_{12}$,$4_{13}$,$6_{10}$,$0_{12}$,$4_{15}$,$4_{16}$,$4_{17}$,$1_{15}$,$3_{13}$,$4_{18}$,$7_{21}$,$4_{19}$,$7_{23}$,$5_{13}$,$0_{14}$ } &  2198 \\
\hline
\end{tabular}
\caption{New Upper Bound solution of Dauzere : 05a. This table provides job and task order and last timestamp for every machine. Job and tasks are formalized as $Job_{Task}$. Solving time is 180000 seconds.}
\end{table}

\begin{table}[h!]
\centering
\begin{tabular}{|c|>{\centering\arraybackslash}p{10cm}|>{\centering\arraybackslash}p{1.5cm}|}
\hline
\textbf{Machine} & \textbf{Job order} & \textbf{end time} \\
\hline
0 & \makecell{ $3_{0}$,$0_{0}$,$6_{0}$,$8_{1}$,$6_{1}$,$1_{3}$,$2_{3}$,$7_{3}$,$8_{5}$ } &  4909 \\
\hline
1 & \makecell{ $1_{0}$,$1_{1}$,$3_{1}$,$3_{2}$,$5_{1}$,$2_{1}$,$7_{1}$,$7_{2}$,$5_{4}$,$8_{4}$,$5_{5}$ } &  4909 \\
\hline
2 & \makecell{ $5_{0}$,$4_{0}$,$0_{1}$,$4_{2}$,$8_{2}$,$9_{2}$,$6_{2}$,$6_{3}$,$1_{4}$,$1_{5}$ } &  4907 \\
\hline
3 & \makecell{ $9_{0}$,$4_{1}$,$9_{1}$,$4_{3}$,$5_{3}$,$8_{3}$,$0_{5}$,$6_{4}$,$6_{5}$,$7_{4}$,$7_{5}$ } &  4909 \\
\hline
4 & \makecell{ $7_{0}$,$3_{3}$,$2_{2}$,$3_{4}$,$9_{3}$,$9_{4}$,$0_{4}$,$4_{4}$,$2_{4}$,$2_{5}$ } &  4909 \\
\hline
5 & \makecell{ $8_{0}$,$2_{0}$,$1_{2}$,$5_{2}$,$0_{2}$,$0_{3}$,$3_{5}$,$9_{5}$,$4_{5}$ } &  4909 \\
\hline
\end{tabular}
\caption{New Upper Bound and optimal solution of Hurink vdata : car5. This table provides job and task order and last timestamp for every machine. Job and tasks are formalized as $Job_{Task}$. Solving time is 1360 seconds}
\end{table}

\begin{table}[h!]
\centering
\begin{tabular}{|c|>{\centering\arraybackslash}p{10cm}|>{\centering\arraybackslash}p{1.5cm}|}
\hline
\textbf{Machine} & \textbf{Job order} & \textbf{end time} \\
\hline
0 &  29, 25, 11, 22, 17, 26, 21, 1, 8, 13, 28, 15, 18, 7, 27, 6, 24, 10, 19, 2, 23, 0, 20, 12, 14, 5, 9, 4, 16, 3 &  1895 \\
\hline
1 &  14, 19, 29, 24, 28, 21, 8, 22, 0, 26, 27, 12, 10, 25, 15, 11, 5, 18, 7, 6, 9, 23, 17, 13, 16, 2, 4, 20, 3, 1 &  \textbf{1997} \\
\hline
2 &  3, 2, 28, 8, 1, 15, 14, 6, 10, 4, 5, 17, 27, 22, 12, 23, 18, 11, 25, 19, 26, 7, 21, 29, 20, 0, 16, 9, 24, 13 &  1995 \\
\hline
3 &  11, 14, 6, 20, 17, 26, 21, 15, 22, 16, 23, 0, 28, 29, 18, 27, 3, 1, 7, 12, 13, 24, 5, 10, 4, 19, 9, 25, 8, 2 &  1974 \\
\hline
4 &  5, 3, 21, 2, 4, 13, 19, 20, 23, 6, 26, 28, 12, 17, 22, 11, 8, 16, 29, 18, 27, 14, 10, 9, 0, 24, 7, 25, 1, 15 &  1822 \\
\hline
5 &  21, 24, 22, 28, 8, 20, 13, 25, 10, 29, 14, 7, 16, 3, 12, 0, 5, 4, 23, 18, 26, 6, 15, 2, 9, 27, 19, 11, 1, 17 &  1985 \\
\hline
6 &  7, 24, 17, 12, 9, 26, 3, 28, 1, 5, 6, 11, 29, 16, 18, 23, 25, 20, 27, 21, 4, 14, 15, 0, 13, 10, 8, 22, 2, 19 &  \textbf{1997} \\
\hline
7 &  17, 15, 21, 28, 11, 23, 22, 10, 27, 18, 25, 20, 2, 5, 1, 9, 6, 29, 0, 3, 14, 4, 19, 8, 7, 16, 12, 26, 24, 13 &  1974 \\
\hline
8 &  24, 1, 28, 21, 3, 17, 25, 18, 14, 11, 12, 6, 7, 10, 20, 27, 5, 9, 26, 19, 15, 4, 16, 23, 0, 22, 8, 13, 2, 29 &  1976 \\
\hline
9 &  11, 7, 3, 17, 18, 26, 24, 6, 12, 23, 0, 16, 4, 9, 25, 28, 15, 5, 20, 1, 22, 14, 10, 29, 13, 8, 19, 2, 21, 27 &  1990 \\
\hline
10 &  8, 20, 3, 10, 28, 29, 12, 15, 25, 23, 0, 6, 21, 11, 2, 1, 9, 5, 27, 18, 13, 14, 4, 16, 26, 24, 7, 19, 22, 17 &  1986 \\
\hline
11 &  28, 0, 6, 26, 10, 13, 9, 21, 23, 14, 1, 15, 4, 17, 11, 27, 22, 20, 18, 25, 19, 24, 3, 5, 2, 29, 16, 7, 12, 8 &  \textbf{1997} \\
\hline
12 &  15, 26, 14, 18, 25, 3, 28, 16, 22, 7, 13, 6, 29, 17, 4, 10, 11, 2, 24, 1, 12, 5, 19, 20, 23, 27, 21, 8, 9, 0 &  \textbf{1997} \\
\hline
13 &  22, 24, 27, 15, 11, 12, 26, 28, 1, 21, 18, 6, 13, 17, 16, 23, 0, 2, 8, 10, 19, 5, 25, 14, 7, 9, 4, 3, 29, 20 &  1858 \\
\hline
14 &  23, 22, 6, 26, 15, 11, 2, 24, 8, 28, 29, 18, 25, 10, 21, 12, 9, 5, 27, 20, 13, 7, 1, 14, 3, 4, 17, 16, 19, 0 &  1969 \\
\hline
15 &  23, 1, 8, 14, 9, 6, 13, 27, 21, 10, 11, 26, 20, 28, 15, 17, 5, 22, 0, 25, 2, 18, 4, 29, 3, 7, 12, 19, 24, 16 &  1904 \\
\hline
16 &  16, 23, 22, 8, 20, 12, 1, 25, 29, 17, 5, 21, 7, 28, 15, 18, 14, 6, 4, 9, 19, 10, 3, 13, 2, 26, 0, 27, 24, 11 &  1995 \\
\hline
17 &  3, 2, 8, 7, 24, 28, 21, 9, 22, 6, 15, 12, 16, 17, 23, 25, 14, 1, 29, 5, 10, 27, 11, 18, 13, 0, 19, 26, 20, 4 &  1996 \\
\hline
18 &  27, 2, 0, 1, 8, 22, 12, 5, 4, 19, 26, 25, 23, 15, 3, 17, 20, 29, 28, 11, 6, 18, 10, 13, 14, 7, 9, 24, 21, 16 &  1981 \\
\hline
19 &  15, 11, 6, 23, 25, 29, 24, 17, 16, 18, 0, 21, 22, 10, 20, 28, 13, 1, 9, 2, 5, 12, 3, 19, 14, 27, 4, 26, 8, 7 &  1990 \\
\hline
\end{tabular}
\caption{Optimal solution of Taillard : ta45. This table provides job order and last timestamp for every machine. Solution is found and said optimal in 154 seconds.}
\end{table}

\begin{table}[h!]
\centering
\begin{tabular}{|c|>{\centering\arraybackslash}p{10cm}|>{\centering\arraybackslash}p{1.5cm}|}
\hline
\textbf{Machine} & \textbf{Job order} & \textbf{end time} \\
\hline
0 &  19, 12, 1, 3, 8, 0, 16, 10, 9, 2, 6, 4, 14, 7, 18, 5, 17, 15, 11, 13 &  1584 \\
\hline
1 &  14, 11, 12, 2, 10, 5, 18, 15, 17, 8, 6, 19, 4, 1, 16, 13, 9, 7, 3, 0 &  1602 \\
\hline
2 &  10, 6, 0, 14, 1, 16, 5, 11, 19, 15, 8, 2, 9, 3, 17, 18, 4, 12, 7, 13 &  1617 \\
\hline
3 &  5, 0, 15, 17, 8, 13, 11, 10, 4, 16, 12, 14, 9, 19, 6, 18, 7, 1, 3, 2 &  1631 \\
\hline
4 &  15, 11, 14, 8, 0, 4, 19, 2, 17, 10, 9, 13, 12, 6, 18, 16, 3, 7, 1, 5 &  1609 \\
\hline
5 &  17, 14, 9, 16, 8, 4, 6, 5, 10, 0, 7, 18, 3, 15, 13, 2, 11, 19, 1, 12 &  1498 \\
\hline
6 &  17, 16, 14, 13, 6, 0, 3, 10, 1, 11, 19, 9, 18, 15, 7, 12, 8, 4, 2, 5 &  1640 \\
\hline
7 &  13, 8, 0, 2, 11, 6, 18, 10, 14, 19, 7, 5, 15, 9, 1, 12, 3, 17, 4, 16 &  1574 \\
\hline
8 &  8, 16, 2, 13, 1, 4, 12, 17, 11, 5, 6, 0, 7, 15, 9, 3, 19, 10, 18, 14 &  1620 \\
\hline
9 &  11, 0, 5, 10, 19, 13, 4, 8, 6, 9, 14, 3, 16, 12, 2, 7, 15, 17, 18, 1 &  \textbf{1643} \\
\hline
10 &  0, 11, 4, 14, 3, 9, 8, 13, 15, 5, 16, 18, 6, 2, 7, 10, 17, 19, 12, 1 &  1409 \\
\hline
11 &  17, 5, 1, 11, 9, 14, 6, 8, 13, 2, 15, 3, 10, 7, 4, 16, 19, 18, 12, 0 &  1482 \\
\hline
12 &  7, 8, 5, 13, 17, 12, 16, 2, 0, 15, 4, 11, 1, 10, 18, 6, 3, 19, 14, 9 &  1594 \\
\hline
13 &  15, 13, 9, 12, 5, 8, 11, 14, 19, 17, 18, 16, 3, 6, 0, 1, 4, 2, 10, 7 &  1640 \\
\hline
14 &  9, 10, 1, 5, 14, 13, 15, 0, 2, 16, 18, 4, 8, 17, 7, 6, 11, 12, 19, 3 &  \textbf{1643} \\
\hline
15 &  16, 3, 11, 10, 0, 2, 19, 14, 1, 17, 8, 15, 12, 6, 9, 7, 4, 18, 13, 5 &  1583 \\
\hline
16 &  16, 12, 4, 19, 13, 6, 3, 11, 18, 2, 8, 5, 17, 1, 10, 14, 9, 15, 0, 7 &  1626 \\
\hline
17 &  9, 10, 13, 12, 17, 0, 5, 4, 8, 2, 7, 11, 14, 16, 3, 18, 19, 1, 15, 6 &  1641 \\
\hline
18 &  15, 4, 0, 11, 2, 3, 14, 18, 7, 17, 6, 10, 1, 12, 5, 8, 9, 13, 19, 16 &  1481 \\
\hline
19 &  11, 17, 16, 5, 15, 6, 12, 7, 8, 3, 13, 18, 19, 14, 4, 1, 0, 10, 9, 2 &  1589 \\
\hline
\end{tabular}
\caption{New Upper Bound solution of Taillard : ta26. This table provides job order and last timestamp for every machine. Solving time is 3600 seconds.}
\end{table}

\begin{table}[h!]
    \centering
    \begin{tabular}{| C{2cm} | C{1.5cm} | C{1.5cm} | }
        \hline
            Instance & LB & UB \\
        \hline 	
            Mk01 & 40 & 40* \\
            Mk02 & 26 & 26* \\
            Mk03 & 204 & 204* \\
            Mk04 & 60 & 60* \\
            Mk05 & 172 & 172* \\            
            Mk06 & 57 & 57* \\
            Mk07 & 139 & 139* \\
            Mk08 & 523 & 523* \\            
            Mk09 & 307 & 307* \\
            Mk10 & \textbf{190} & 193 \\
        \hline
    \end{tabular}
    \caption{
    Consolidated bounds on Brandimarte dataset \cite{brandimarte1993routing}. New values are in \textbf{bold}. Solution is optimal when UB equal LB, which is identified with *}
\end{table}

\begin{table}[h!]
    \centering
    \begin{tabular}{| C{2cm} | C{1.5cm} | C{1.5cm} | C{0.2cm} |  C{2cm} | C{1.5cm} | C{1.5cm} |}
        \hline
            Instance & LB & UB & & Instance & LB & UB \\
        \hline 	
            Ta01 & 1231 & 1231* & & Ta41 & \textbf{1915} & 2006 \\
            Ta02 & 1244 & 1244* & & Ta42 & \textbf{1888} & 1939 \\
            Ta03 & 1218 & 1218* & & Ta43 & 1809 & 1846 \\
            Ta04 & 1175 & 1175* & & Ta44 & 1952 & 1979 \\
            Ta05 & 1224 & 1224* & & Ta45 & 1997 & \textbf{1997*} \\
            Ta06 & 1238 & 1238* & & Ta46 & \textbf{1967} & 2006 \\
            Ta07 & 1227 & 1227* & & Ta47 & \textbf{1819} & 1889 \\
            Ta08 & 1217 & 1217* & & Ta48 & 1915 & 1937 \\
            Ta09 & 1274 & 1274* & & Ta49 & 1934 & 1960 \\
            Ta10 & 1241 & 1241* & & Ta50 & 1837 & 1923 \\
            Ta11 & 1357 & 1357* & & Ta51 & 2760 & 2760* \\
            Ta12 & 1367 & 1367* & & Ta52 & 2756 & 2756* \\
            Ta13 & 1342 & 1342* & & Ta53 & 2717 & 2717* \\
            Ta14 & 1345 & 1345* & & Ta54 & 2839 & 2839* \\
            Ta15 & 1339 & 1339* & & Ta55 & 2679 & 2679* \\
            Ta16 & 1360 & 1360* & & Ta56 & 2781 & 2781* \\
            Ta17 & 1462 & 1462* & & Ta57 & 2943 & 2943* \\
            Ta18 & \textbf{1393} & 1396 & & Ta58 & 2885 & 2885* \\
            Ta19 & 1332 & 1332* & & Ta59 & 2655 & 2655* \\
            Ta20 & 1348 & 1348* & & Ta60 & 2723 & 2723* \\
            Ta21 & 1642 & 1642* & & Ta61 & 2868 & 2868* \\
            Ta22 & \textbf{1581} & 1600 & & Ta62 & 2869 & 2869* \\
            Ta23 & \textbf{1544} & 1557 & & Ta63 & 2755 & 2755* \\
            Ta24 & 1644 & 1644* & & Ta64 & 2702 & 2702* \\
            Ta25 & \textbf{1586} & 1595 & & Ta65 & 2725 & 2725* \\
            Ta26 & \textbf{1616} & \textbf{1643} & & Ta66 & 2845 & 2845* \\
            Ta27 & \textbf{1673} & 1680 & & Ta67 & 2825 & 2825* \\
            Ta28 & 1603 & 1603* & & Ta68 & 2784 & 2784* \\
            Ta29 & \textbf{1607} & 1625 & & Ta69 & 3071 & 3071* \\
            Ta30 & \textbf{1552} & 1584 & & Ta70 & 2995 & 2995* \\
            Ta31 & 1764 & 1764* & & Ta71 & 5464 & 5464* \\
            Ta32 & 1774 & 1784 & & Ta72 & 5181 & 5181* \\
            Ta33 & \textbf{1791} & 1791\textbf{*} & & Ta73 & 5568 & 5568* \\
            Ta34 & 1828 & 1828* & & Ta74 & 5339 & 5339* \\
            Ta35 & 2007 & 2007* & & Ta75 & 5392 & 5392* \\
            Ta36 & 1819 & 1819* & & Ta76 & 5342 & 5342* \\
            Ta37 & 1771 & 1771* & & Ta77 & 5436 & 5436* \\
            Ta38 & 1673 & 1673* & & Ta78 & 5394 & 5394* \\
            Ta39 & 1795 & 1795* & & Ta79 & 5358 & 5358* \\
            Ta40 & \textbf{1653} & 1670 & & Ta80 & 5183 & 5183* \\

        \hline
    \end{tabular}
    \caption{Consolidated bounds on Taillard dataset \cite{taillard1993benchmarks}. New values are in \textbf{bold}. Solution is optimal when UB equal LB, which is identified with *}
\end{table}

\begin{table}[h!]
    \centering
    \begin{tabular}{| C{2cm} | C{1.5cm} | C{1.5cm} | }
        \hline
            Instance & LB & UB \\
        \hline 	
            01a & 2505 & 2505* \\
            02a & 2228 & 2228* \\
            03a & 2228 & 2228* \\
            04a & 2503 & 2503* \\
            05a & \textbf{2195} & \textbf{2199} \\
            06a & \textbf{2164} & \textbf{2169} \\
            07a & 2216 & 2254 \\
            08a & 2061 & 2061* \\
            09a & 2061 & 2061* \\
            10a & 2212 & 2241 \\
            11a & \textbf{2020} & 2037 \\
            12a & 1969 & 1984 \\
            13a & \textbf{2206} & 2236 \\
            14a & 2161 & 2161* \\
            15a & 2161 & 2161* \\
            16a & \textbf{2202} & 2231 \\
            17a & \textbf{2089} & 2105 \\
            18a & 2057 & 2070 \\
        \hline
    \end{tabular}
    \caption{Consolidated bounds on Dauzere dataset \cite{dauzere1995solving}. New values are in \textbf{bold}. Solution is optimal when UB equal LB, which is identified with *}
\end{table}

\begin{table}[h!]
    \centering
    \begin{tabular}{| C{3cm} | C{1.5cm} | C{1.5cm} | C{0.2cm} |  C{3cm} | C{1.5cm} | C{1.5cm} |}
        \hline
            Instance & LB & UB & & Instance & LB & UB \\
        \hline 	
            cscmax\_20\_15\_1 & 3181 & 3266 &  & rcmax\_20\_15\_1 & 2749 & 2749* \\
            cscmax\_20\_15\_10 & \textbf{3139} & 3248 &  & rcmax\_20\_15\_10 & 2706 & 2706* \\
            cscmax\_20\_15\_5 & 3321 & 3390 &  & rcmax\_20\_15\_4 & 2563 & 2563* \\
            cscmax\_20\_15\_7 & 3386 & 3475 &  & rcmax\_20\_15\_5 & 2731 & 2731* \\
            cscmax\_20\_15\_8 & 3390 & 3441 &  & rcmax\_20\_15\_8 & \textbf{2641} & 2669 \\
            cscmax\_20\_20\_2 & 3515 & 3706 &  & rcmax\_20\_20\_4 & \textbf{3014} & 3046 \\
            cscmax\_20\_20\_3 & 3583 & 3763 &  & rcmax\_20\_20\_5 & 2984 & 2984* \\
            cscmax\_20\_20\_4 & 3684 & 3939 &  & rcmax\_20\_20\_6 & \textbf{3207} & 3244 \\
            cscmax\_20\_20\_6 & 3740 & 4035 &  & rcmax\_20\_20\_7 & 3188 & 3188* \\
            cscmax\_20\_20\_9 & 3588 & 3729 &  & rcmax\_20\_20\_8 & 3092 & 3092* \\
            cscmax\_30\_15\_10 & 4234 & 4378 &  & rcmax\_30\_15\_1 & 3343 & 3343* \\
            cscmax\_30\_15\_2 & \textbf{4053} & 4156 &  & rcmax\_30\_15\_10 & \textbf{3481} & 3492 \\
            cscmax\_30\_15\_5 & 4271 & 4361 &  & rcmax\_30\_15\_4 & 3394 & 3394* \\
            cscmax\_30\_15\_6 & 4180 & 4258 &  & rcmax\_30\_15\_5 & 3681 & 3681* \\
            cscmax\_30\_15\_9 & \textbf{4179} & 4297 &  & rcmax\_30\_15\_9 & \textbf{3397} & 3430 \\
            cscmax\_30\_20\_3 & 4449 & 4701 &  & rcmax\_30\_20\_10 & \textbf{3723} & 3812 \\
            cscmax\_30\_20\_4 & 4447 & 4721 &  & rcmax\_30\_20\_2 & 3619 & 3699 \\
            cscmax\_30\_20\_6 & 4346 & 4607 &  & rcmax\_30\_20\_7 & 3734 & 3750 \\
            cscmax\_30\_20\_7 & 4444 & 4647 &  & rcmax\_30\_20\_8 & 3697 & 3764 \\
            cscmax\_30\_20\_9 & 4729 & 4939 &  & rcmax\_30\_20\_9 & 3844 & 3844* \\
            cscmax\_40\_15\_3 & \textbf{5016} & 5169 &  & rcmax\_40\_15\_10 & 4668 & 4668* \\
            cscmax\_40\_15\_4 & 5148 & 5226 &  & rcmax\_40\_15\_2 & 4164 & 4164* \\
            cscmax\_40\_15\_6 & 5160 & 5247 &  & rcmax\_40\_15\_5 & 4380 & 4380* \\
            cscmax\_40\_15\_7 & \textbf{5122} & 5173 &  & rcmax\_40\_15\_8 & 4648 & 4648* \\
            cscmax\_40\_15\_8 & 5232 & 5312 &  & rcmax\_40\_15\_9 & 4725 & 4725* \\
            cscmax\_40\_20\_10 & 5512 & 5701 &  & rcmax\_40\_20\_1 & 4647 & 4647* \\
            cscmax\_40\_20\_5 & \textbf{5502} & 5688 &  & rcmax\_40\_20\_2 & 4691 & 4691* \\
            cscmax\_40\_20\_6 & \textbf{5652} & 5779 &  & rcmax\_40\_20\_3 & 4848 & 4848* \\
            cscmax\_40\_20\_8 & 5493 & 5763 &  & rcmax\_40\_20\_6 & 4692 & 4692* \\
            cscmax\_40\_20\_9 & \textbf{5621} & 5868 &  & rcmax\_40\_20\_7 & 4732 & 4732* \\
            cscmax\_50\_15\_10 & \textbf{6107} & 6136 &  & rcmax\_50\_15\_1 & 5927 & 5927* \\
            cscmax\_50\_15\_3 & \textbf{6123} & 6189 &  & rcmax\_50\_15\_2 & 5728 & 5728* \\
            cscmax\_50\_15\_4 & \textbf{6168} & 6196 &  & rcmax\_50\_15\_3 & 5640 & 5640* \\
            cscmax\_50\_15\_6 & \textbf{6434} & 6463 &  & rcmax\_50\_15\_4 & 5385 & 5385* \\
            cscmax\_50\_15\_8 & \textbf{6127} & 6207 &  & rcmax\_50\_15\_5 & 5635 & 5635* \\
            cscmax\_50\_20\_1 & \textbf{6468} & 6718 &  & rcmax\_50\_20\_2 & 5621 & 5621* \\
            cscmax\_50\_20\_3 & 6638 & 6755 &  & rcmax\_50\_20\_3 & 5577 & 5577* \\
            cscmax\_50\_20\_4 & \textbf{6518} & 6747 &  & rcmax\_50\_20\_6 & 5713 & 5713* \\
            cscmax\_50\_20\_7 & 6709 & 6910 &  & rcmax\_50\_20\_7 & 5851 & 5851* \\
            cscmax\_50\_20\_9 & 6459 & 6634 &  & rcmax\_50\_20\_9 & 5747 & 5747* \\

        \hline
    \end{tabular}
    \caption{Consolidated bounds on Demirkol dataset \cite{demirkol1997computational}. New values are in \textbf{bold}. Solution is optimal when UB equal LB, which is identified with *}
\end{table}

\end{document}